\documentclass[11pt]{article} 
\usepackage{mystyle-new}
\usepackage{authblk}
\usepackage[T1]{fontenc}
\usepackage{epsfig,amsmath} 
\usepackage{hepnames,hepunits}
\usepackage{color}
\usepackage{graphicx}
\definecolor{red}{rgb}{1,0,0}
\def\lesssim{\ \hbox{\raise 2pt \hbox{$<$} \kern -13pt
                     \lower 3pt \hbox{$\sim$}}\ }
\def\greatersim{\ \hbox{\raise 2pt \hbox{$>$} \kern -13pt
                     \lower 3pt \hbox{$\sim$}}\ }

\def\lsim{\mathrel{\rlap{\lower4pt\hbox{\hskip1pt$\sim$}}
    \raise1pt\hbox{$<$}}}                
\def\gsim{\mathrel{\rlap{\lower4pt\hbox{\hskip1pt$\sim$}}
    \raise1pt\hbox{$>$}}}                

\newenvironment{tolerant}[1]{\par\tolerance=#1\relax}{ \par }
\usepackage{amsmath,bm}
\usepackage{lineno}

\usepackage{cite,mcite}
\usepackage{tikz}
\usepackage[symbol]{footmisc}

\providecommand{\DOI}[1]{\href{http://dx.doi.org/#1}}

\usepackage{wrapfig}
\usepackage{orcidlink}

\begin{document}


\title{\LARGE Science4Peace: \\
A Plea for Continued Peaceful International Scientific Cooperation 
\\
\vspace{3mm}
\large \bf Input to the European Strategy for Particle Physics - 2026 update   \\
\vspace{3mm}
\large (On behalf of the Science4Peace Forum)}

\author[1]{
A.~Ali~\orcidlink{0000-0002-1939-1545}, 
M.~Barone~\orcidlink{0000-0002-2115-4055}, 
D.~Britzger~\orcidlink{0000-0002-9246-7366},
A.~Cooper-Sarkar~\orcidlink{0000-0002-7107-5902},
J.~Ellis~\orcidlink{0000-0002-7399-0813},
S.~Franchoo~\orcidlink{0000-0001-7520-5922},
A.~Giammanco~\orcidlink{0000-0001-9640-8294},
A.~Glazov~\orcidlink{0000-0002-8553-7338}, 
H.~Jung \thanks{Corresponding author and contact person: hannesjung@science4peace.com}~\orcidlink{0000-0002-2964-9845}, 
D.~K\"afer~\orcidlink{0000-0003-0998-2023},
J.~List~\orcidlink{0000-0002-0626-3093}, 
L.~L\"onnblad~\orcidlink{0000-0003-1269-1649},
M.~Mangano~\orcidlink{0000-0002-0886-3789},
N.~Raicevic~\orcidlink{0000-0002-2386-2290}, 
A. Rostovtsev~\orcidlink{0000-0001-9906-0764},
M.~Schmelling~\orcidlink{0000-0003-3305-0576}
T.~Sch\"ucker~\orcidlink{0000-0001-5551-0444},
A.~Tanasijczuk~\orcidlink{0009-0008-7445-8002},
P.~Van~Mechelen~\orcidlink{0000-0002-8731-9051}
}

\begin{titlepage} 
\maketitle

\begin{abstract}
\vskip 0.5cm
The European Strategy for Particle Physics (ESPP) - 2026 update is taking place in a turbulent international climate. Many of the norms that have governed relations between states for decades are being broken or challenged. The future progress of science in general, and particle physics in particular, will depend on our ability to maintain peaceful international scientific collaboration in the face of political pressures. We plead that the ESPP 2026 update acknowledge explicitly the importance of peaceful international scientific collaboration, not only for the progress of science, but also as a precious bridge between  geopolitical blocs.\\
~~\\
{\it Scientific thought is the common heritage of mankind} - Abdus Salam\\
\end{abstract}

\end{titlepage}

\section{Science4Peace: A Plea for Scientific Diplomacy}
\label{sec:Intro}

Fundamental science is a universal human endeavour, of interest to all, and potentially of benefit to all. As such, it should be free of geopolitical restrictions on participation in its research projects and on access to its results.

In the face of global crises and geopolitical pressures, the field of particle physics - devoted to the peaceful pursuit of fundamental knowledge - stands at a crossroads and is exposed to international tensions. The rising tide of 
restrictions on international collaboration and academic freedom pose threats to the very foundations of the scientific enterprise. Recent developments, including strained relations between major scientific powers such as Europe, the United States, Russia and China, as well as ongoing conflicts in the Middle East, have led to political pressures to exclude researchers based on nationality, institutional or political affiliation. Moreover, climate change, global energy crisis,  global pollution, and the militarization of research raise pressing ethical questions about the future of our field.

Scientific collaboration has played a vital role in particle physics and bridging political and ideological divides~\cite{Heuer:2015eca,CERNcourier-Schopper,VanMechelen:2023cer,Albrecht:2023xxi,Ali:2024fxg,opinion-vie-CERN-courier}. CERN was established in 1954 to unite nations through peaceful scientific research~\cite{cern_convention_1954}, fostering cooperation among former adversaries in the wake of World War II. Similarly, the Joint Institute for Nuclear Research\cite{JINR-Charter} (JINR) was founded in Dubna in 1956 as a symbol of international collaboration in fundamental physics, bringing together scientists from across political divides. The HERA accelerator at DESY~\cite{DESY-guiding_principles} further exemplified this spirit by bringing together researchers from the United States, USSR, East and West Germany, and many other nations at both sides of the iron curtain, demonstrating the power of scientific unity even during periods of geopolitical tension. The LHC at CERN has brought together international collaborations that are unprecedented in size and in the numbers of national teams, often coming from countries with tense bilateral relations.  Other examples of bridges across political and ideological divides are the SESAME project~\cite{SESAME-unesco,SESAME-home} in the Middle East, which brings together countries with tense relations, as well as for the proposal to build  a similar scientific infrastructure in the Western Balkans called SEEIIST~\cite{SEEIST-home}, which brings together scientists from  Albania, Bosnia and Herzegovina, Bulgaria, Greece, Kosovo, Montenegro, North Macedonia, Serbia and Slovenia.
We note also that open access international conferences, workshops and schools play a crucial role for communication across political and ideological divides.

Scientists in democratic societies are coming under increasing political pressure, restrictions on international travel and cooperation, as well as funding cuts and limitations on academic freedom~\cite{Kinzelbach2025}, problems long familiar to their colleagues in other regimes. 
Moreover, the integrity of scientific inquiry is being challenged by competition between geopolitical blocs, barriers to open-access science and the increasing commodification of research. As members of the international particle physics community, we recognize our responsibility to advocate for ethical principles, peace, and sustainability in research~\cite{Bloom:2022gux,Banerjee:2023avd}. We call upon the members of the HEP community,  institutions, policymakers, and funding bodies to uphold the following principles:

\section{Guiding Principles for Science for Peace}  
\label{sec:RecommendationS4P}

\begin{itemize}

\item Commitment to Peaceful Civilian Scientific Research
\begin{itemize}
\item As set out in the CERN convention~\cite{cern_convention_1954} and the DESY Mission statement~\cite{DESY-guiding_principles}, research in particle physics should refrain from military applications, and institutions should refuse funding or partnerships linked to military objectives.

\end{itemize}

\item International Collaboration and Inclusion
\begin{itemize}
\item Scientific collaboration should transcend political and national divides~\cite{CERNcourier-Schopper}, in accordance with the founding principles of CERN, DESY and JINR.

\item Scientists should not be excluded from international projects on the basis of their nationality or institutional affiliation as long as they agree to conduct only peaceful civilian research.

\item Scientific collaboration should follow the principles of diversity, equity and inclusion.

\item Measures should be taken to extend research opportunities to scientists working under challenging conditions.

\end{itemize}

\end{itemize}

\section{Recommendations for Promoting Peaceful International Scientific Cooperation}

\begin{itemize}

\item Funding for Peaceful Civilian Collaboration
\begin{itemize}

\item Science should be a peaceful civilian collaborative effort for the benefit of humanity and not a competition between rival geopolitical blocs.

\item International funding initiatives to support international and inter-regional research efforts should be strengthened.

\item Funding for fundamental research institutions should be free from geopolitical and market-driven constraints.

\end{itemize}

\item Open Science and Free Access to Knowledge
\begin{itemize}

\item All publicly funded research should be openly accessible to the global scientific community and to humankind in general, in accordance with the UN declaration of human rights, art.\ 27~\cite{un_udhr_1948}).

\item Research institutions should promote transparency, data-sharing, and open-access publication models.

\item Conferences, workshops and schools should  be held without access restrictions

\item Outreach efforts should be strengthened, so as to ensure free and equitable access to scientific knowledge for all, to counter misinformation and disinformation, and to encourage the resolution of problems through evidence and logic in an environment of mutual respect.

\end{itemize}

\item Sustainability and Ethical Research Practices
\begin{itemize}

\item Research institutions should adopt sustainable practices to minimize the environmental impact of large-scale physics experiments.

\item Efforts must be made to reduce carbon emissions, manage waste responsibly, and promote sustainable energy use.

\item Research institutions and funding agencies should follow the basic principles of universal participation in, and access to, the results of the research programmes that they support.

\end{itemize}

\item Protection of Academic Freedom and Freedom of Speech
\begin{itemize}

\item Governments and institutions must protect the freedom of scientists to conduct independent research and express their views without political or ideological interference, in accordance with EU Charter Art. 13~\cite{EUCharter_Article13}).

\item National policies should align with UN and EU principles on academic freedom and free speech.

\end{itemize}
\end{itemize}

\section{Conclusion}

We urge the high-energy physics community, its research institutions, funding agencies and national governments to uphold the principles of peaceful international collaboration, in order to ensure that science remains a force for peace, progress, and the betterment of society. The international particle physics community must stand united in the face of global challenges, defending ethical values, freedom, and collaboration for a better future.

\vskip 0.5 cm 
\begin{tolerant}{8000} 
\noindent 
{\bf Acknowledgments.} 
We thank all participants of the {\it Science4Peace Forum} for supportive discussions. 

\noindent 

\end{tolerant} 

\bibliographystyle{mybibstyle-new}
\raggedright  
\providecommand{\href}[2]{#2}\begingroup\raggedright\endgroup


\begin{thebibliography}{10}%
\makeatletter
\providecommand{\hrefCMSnoop }[0]{\@secondoftwo}%
\makeatother
\providecommand{\doi}{\texttt{doi:}\begingroup \urlstyle{tt}\Url}

\bibitem{Heuer:2015eca}
\hrefCMSnoop {}{R.-D. Heuer, ``{CERN and 60 years of science for peace}'',}
  \textit{ AIP Conf. Proc.} \textbf{ 1645} (2015), no.~1, 430--436.

\bibitem{CERNcourier-Schopper}
\href
  {https://cerncourier.com/a/science-for-peace-more-than-ever/}{H.~Schopper,
  ``Science For Peace? More than ever!'',} \textit{ CERN Courier} \textbf{ 62}
  (2022), no.~5, 49.

\bibitem{VanMechelen:2023cer}
\hrefCMSnoop {}{P.~Van~Mechelen, ``{Science4Peace in difficult times}'',}
  \textit{ PoS} \textbf{ EPS-HEP2023} (2024) 650,
  \href{http://www.arXiv.org/abs/2311.05484}{\texttt{arXiv:2311.05484}}.

\bibitem{Albrecht:2023xxi}
\href {https://arxiv.org/abs/2311.02141}{M.~Albrecht {et~al.}, ``{Beyond a Year
  of Sanctions in Science}'',}
\newblock (2023).
\newblock
  \href{http://www.arXiv.org/abs/2311.02141}{\texttt{arXiv:2311.02141}}.

\bibitem{Ali:2024fxg}
\href {https://arxiv.org/abs/2403.07833}{A.~Ali {et~al.}, ``{A Science4Peace
  initiative: Alleviating the consequences of sanctions in international
  scientific cooperation}'',}
  \href{http://www.arXiv.org/abs/2403.07833}{\texttt{arXiv:2403.07833}}.

\bibitem{opinion-vie-CERN-courier}
\href
  {https://cerncourier.com/a/science-needs-cooperation-not-exclusion/}{H.~Jung,
  ``Science needs cooperation, not exclusion''.} CERN Courier, 2024.

\bibitem{cern_convention_1954}
\href {https://cds.cern.ch/record/330625}{CERN, ``Convention for the
  Establishment of a European Organization for Nuclear Research, Article II'',}
  (1954).

\bibitem{JINR-Charter}
\href
  {https://www.jinr.ru/wp-content/uploads/Advisory_Bodies/JINR_Charter_1992_eng.pdf}{{JINR},
  ``Charter of the Joint Institute for Nuclear Research'',} (1992).

\bibitem{DESY-guiding_principles}
\href
  {https://www.desy.de/about_desy/mission_and_guiding_principles/index_eng.html}{\mbox{DESY},
  ``Mission and guiding principles of DESY'',} (2013).

\bibitem{SESAME-unesco}
\href
  {https://en.unesco.org/courier/2018-4/sesame-scientific-excellence-middle-east}{``SESAME:
  scientific excellence in the Middle East''.}

\bibitem{SESAME-home}
\href {https://www.sesame.org.jo/}{``Synchrotron-light for Experimental Science
  and Applications in the Middle East (SESAME)''.}

\bibitem{SEEIST-home}
\href {https://seeiist.eu/}{``The South East European International Institute
  for Sustainable Technologies (SEEIIST)''.}

\bibitem{Kinzelbach2025}
\hrefCMSnoop {}{K.~Kinzelbach, S.~I. Lindberg, L.~Lott, and A.~V. Panaro,
  ``Academic Freedom Index Update 2025''.}
  https://doi.org/10.25593/open-fau-1637, (2025).

\bibitem{Bloom:2022gux}
K.~Bloom\hrefCMSnoop {}{ {et~al.}, ``{Climate impacts of particle physics}'',}
  in \textit{ {2022 Snowmass Summer Study}}.
\newblock 3, (2022).
\newblock
  \href{http://www.arXiv.org/abs/2203.12389}{\texttt{arXiv:2203.12389}}.

\bibitem{Banerjee:2023avd}
\hrefCMSnoop {}{S.~Banerjee {et~al.}, ``{Environmental sustainability in basic
  research. A perspective from HECAP+}'',} \textit{ JINST} \textbf{ 20} (2025)
  P03012,
  \href{http://www.arXiv.org/abs/2306.02837}{\texttt{arXiv:2306.02837}}.

\bibitem{un_udhr_1948}
\hrefCMSnoop {}{{United Nations}, ``Universal Declaration of Human Rights,
  Article 27'',}
  https://www.un.org/en/about-us/universal-declaration-of-human-rights, (1948).

\bibitem{EUCharter_Article13}
\href
  {https://eur-lex.europa.eu/legal-content/EN/TXT/?uri=CELEX%3A12012P013}{{European
  Union}, ``{Charter of Fundamental Rights of the European Union, Article 13 -
  Freedom of the Arts and Sciences}'',} (2012).
\newblock Official Journal of the European Union, C 326/391.

\end{thebibliography}

\end{document}